\documentclass[conference]{basi}
\usepackage[british]{babel}
\usepackage[varg]{txfonts}
\usepackage{rotating}
\usepackage{dcolumn}
\begin{document}
\title[Gas and Dust in the MCs]{Gas and Dust in the Magellanic Clouds}
\author[Pradhan Ananta C.]%
       {Ananta~C.~Pradhan\thanks{email: \texttt{acp@tifr.res.in}}\\
Tata Institute of Fundamental Research, Homi Bhabha Road, Colaba,
Mumbai 400 005, India}



\date{}
\maketitle
\label{firstpage}

\begin{abstract}
 The far-ultraviolet (FUV) diffuse emission is predominantly due to scattering of starlight from interstellar dust grains which shows a large regional variation depending on the relative orientations of dust and stars. The observations of the FUV (1000 -- 1150 \AA) diffuse radiation in the Magellanic Clouds (MCs) using serendipitous observations made with the Far Ultraviolet {\em Spectroscopic Explorer (FUSE)} are presented. The estimated contribution of FUV diffuse radiation to the total integrated FUV emission in the MCs is found to be typically 5\% -- 20\% in the Large Magellanic Cloud (LMC) and 34\% -- 44\% in the Small Magellanic Cloud (SMC) at the {\em FUSE} bands ($\lambda$ = 905 -- 1187 \AA) and it increases substantially towards the longer wavelength (e.g., 63\% for the SMC at 1615 \AA). The less scattering of light in the FUV at the shorter wavelength than at the longer wavelength indicates that much of the stellar radiation at the shorter wavelength is going into heating the interstellar dust.

Five times ionized oxygen atom (O {\small VI}) is a tracer of hot gas (T $\sim 3\times 10^{5}$ K) in the interstellar medium (ISM). A wide survey of O {\small VI} column density measurements for the LMC is presented using the high resolution {\em FUSE} spectra. The column density varies from a minimum of log N(O {\small VI}) = 13.72 atoms cm$^{-2}$ to a maximum of log N(O {\small VI}) = 14.57 atoms cm$^{-2}$. A high abundance of O {\small VI} is observed in both active (superbubbles) and inactive regions of the LMC.

\end{abstract}


\section{Introduction}

The MCs are proximate extragalactic objects which provide an unique oppertunity to study the spatial and temporal distribution of ISM. Dust in the MCs is known to be different from the dust in the Milky Way which is attributed to low metalicity, low dust to gas ratio and different behavior of extinction of them. The MCs are oriented nearly face on view with low foreground extinction which allows the observer to study the interstellar emission processes along the line of sight without any confusion. Apart from these, the ISM of the MCs can be treated as primitive and therefore, may be the stepping stone to our understanding of the ISM in high red shift galaxies. Diffuse UV background light in the MCs is from hot stars scattered by interstellar dust grains. The intensity of this radiation depends on the distribution of hot stars, amount and distribution of dust grains, optical depth and the optical parameters of dust.

O {\small VI} which has a doublet of transitions at 1032 \AA\ and 1038 \AA\ is a diagnostic of hot gas in the ISM. This is collisionally ionized and produced at the interface of hot (T $> 10^{6}$ K) and warm (T $\sim 10^{4}$ K) ionized gas. The FUSE, with a spectral resolution of $\sim$ 20,000 provided a detailed study of O VI absorption and emission in the ISM of the Milky Way and the MCs as well as in the Intergalactic medium. 

\section{Result and Discussion}
\begin{figure}
\centerline{\includegraphics[width=7.0cm]{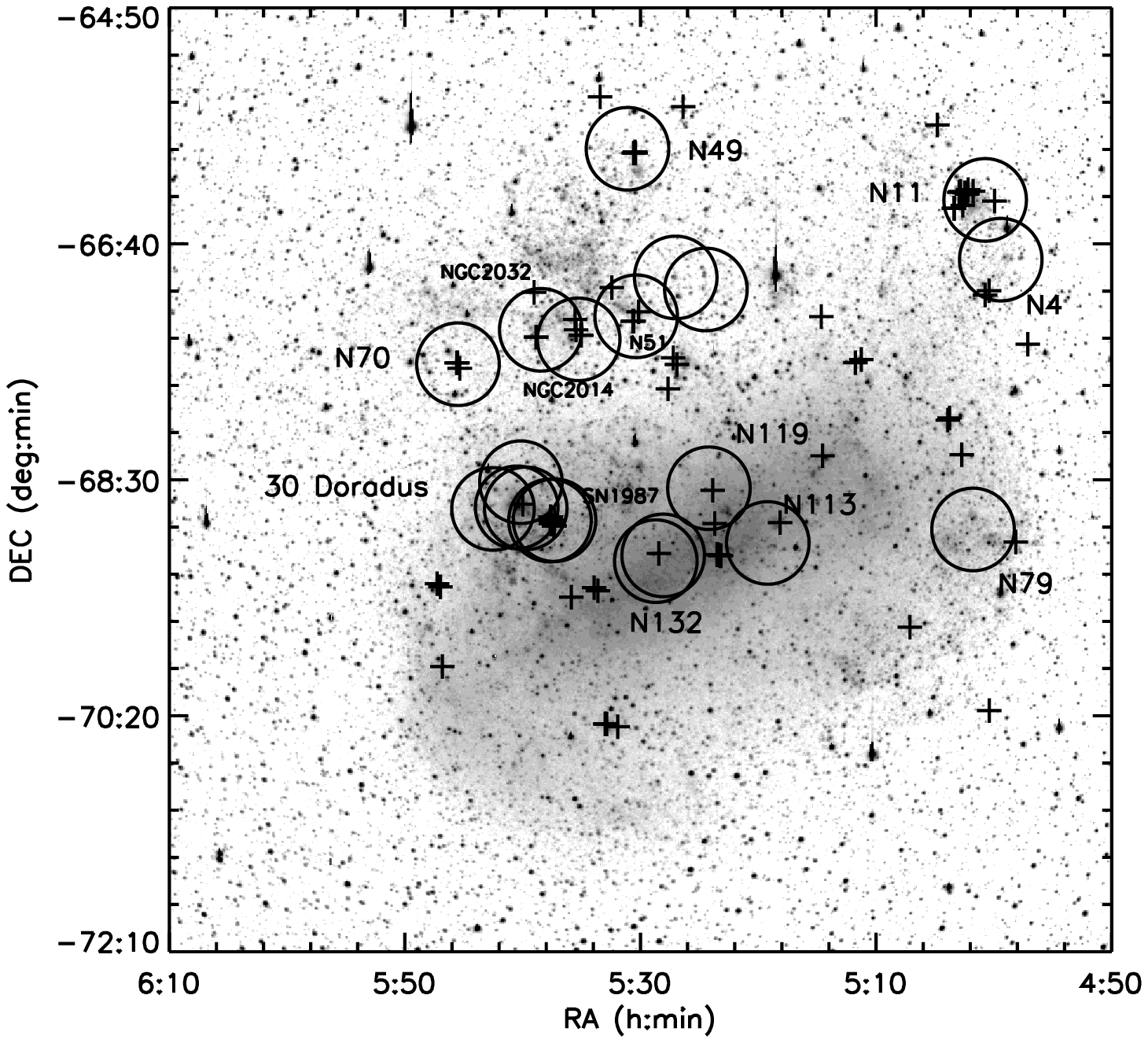} \qquad
            \includegraphics[width=6.0cm]{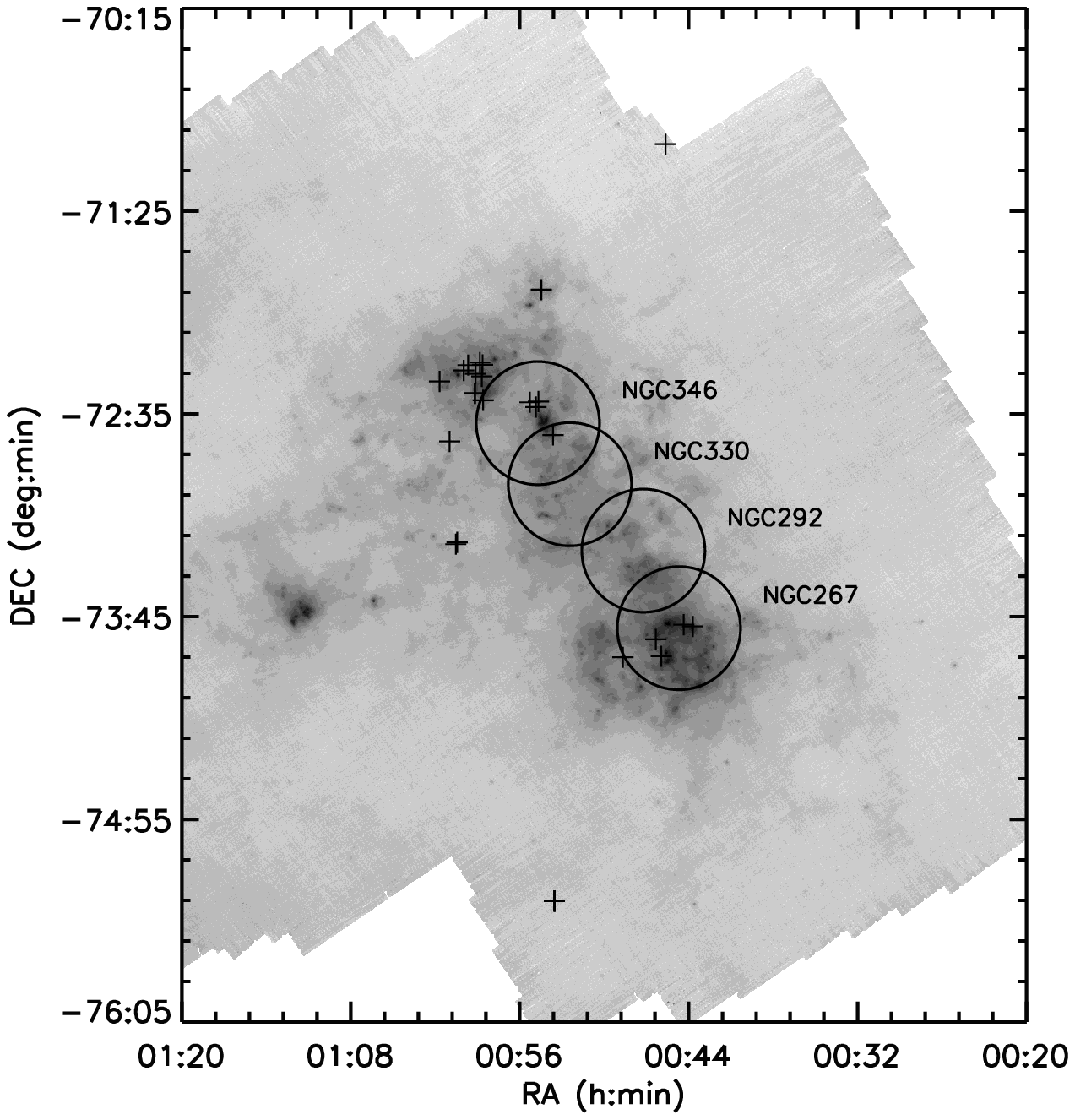}}
\caption{{\em FUSE} observations are represented by `+' symbols and the UIT observations are shown by black circles in the maps of the MCs ({\it left panel:} R-band image of the LMC from Bothun \& Thumpson (1988); {\it Right panel:} IR 160$\mu$m image of the SMC from Gordon et al. (2009))}
\end{figure}
All the archival data of {\em FUSE} in the MCs were examined for the suitability of diffuse measurements using the data analysis method of Murthy \& Sahnow (2004). Finally, 81 out of 600 pointings in the LMC and 30 out of 220 pointings in the SMC were found to be suitable diffuse observations as shown in Figure 1. Most of these observations were near important regions of the MCs (Pradhan, Pathak \& Murthy 2010; Pradhan, Murthy \& Pathak 2011). The intensities of these observations range from around $10^{3}$ photons cm$^{-2}$ s$^{-1}$ sr$^{-1}$ \AA$^{-1}$ to as high as $3 \times 10^{5}$ photons cm$^{-2}$ s$^{-1}$ sr$^{-1}$ \AA$^{-1}$ at {\em FUSE} bands with a negligibly small contribution from the Galactic diffuse radiation. 

Some of the diffuse {\em FUSE} observations were within the \(37'\) {\em Ultraviolet Imaging telscope (UIT)} fields in the MCs for which the {\em UIT} diffuse surface brightness were measured. Using these diffuse observations along with the stellar catalogs provided by Parker et al. (1998) for the LMC and Cornett et al. (1997) for the SMC, the fraction of the diffuse emission (diffuse/(diffuse+stellar)) emanating from the {\em UIT} fields at the {\em FUSE} and {\em UIT} bands were estimated. The {\em UIT} fields used for the estimation of FUV diffuse fraction in the LMC are: N70, SN 1987 N51, NGC 2014, N113, N119, N132, N4, N11 and 30 Doradus. In the LMC, the diffuse fraction ranges from 5\% to 20\% at 1100 \AA, with a high of 45\% in the superbubble N70, and with an observed error of 12\% -- 17\%. Integrating over the entire {\em UIT} fields of the LMC, it was found that $\sim$5\% -- 20\% of the total emission is escaping out of the LMC as the diffuse emission. 

Similarly, the {\em UIT} fields in the SMC used for the calculation of FUV diffuse fraction are: NGC 346, NGC 330, NGC 292 and NGC 267. The diffuse emission coming out of these regions shows a large regional variation due to their defference in the stellar contents. The diffuse FUV fraction escaping out of the SMC ranges between 34\% and 44\% in the {em FUSE} bands (1000 -- 1150 \AA) with a further increase upto 63\% at 1615 \AA. This is in consistent with the model calculation of Witt \& Gordon (2000) which was $\sim$25\% - 50\% of the total emission depending on different dust geometries. The diffuse fraction is much less at the {\em FUSE} bands than the corresponding fraction emitted at 1500 \AA\ suggesting that the largest part of the heating of the interstellar dust occurs in the FUV. The albedo of the dust obtained from the theoretical predictions of Weingartner \& Draine (2001) for a mix of spherical carbonaceous and silicate grains shows similar trend like the FUV diffuse fraction. The diffuse fraction in the SMC was found to be higher than the LMC which might be due to the relatively higher value of the albedo of the SMC dust compared to the LMC dust (Pradhan, Murthy \& Pathak 2011). 

\begin{figure}
\centerline{\includegraphics[width=8.0cm]{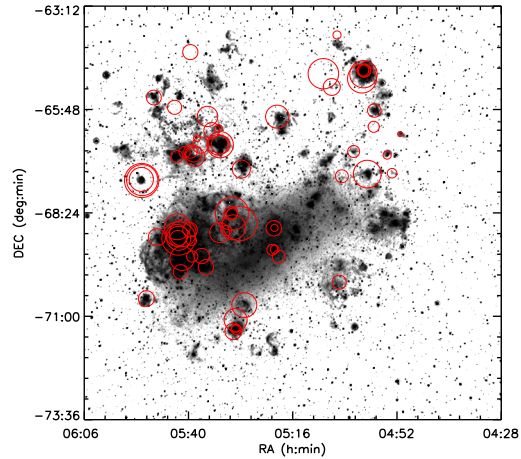}}
\caption{H$\alpha$ image of the LMC with circles representing O VI absorption around the 70 targets. The diameter of the circle is linearly proportional to the column density of O {\small VI} (Pathak et al. 2011).}
\end{figure}
 A wide coverage of O {\small VI} absorption for the LMC is presented using the high resolution absorption spectra of {\em FUSE} for 70 lines of sight. The measurement of O {\small VI} column density is done following the apparent optical depth technique (Howk et al. 2002, Savage \& Sembach 1991, Sembach \& Savage 1992). The O {\small VI} column density in the LMC varies from a minimum of log N(O {\small VI}) = 13.72 atoms cm$^{-2}$ to a maximum value of log N(O~{\small VI}) = 14.57 atoms cm$^{-2}$. The location of the O {\small VI} sight lines are displayed by circles in Figure 2 and the diameter of the circles is proportional to the O {\small VI} column density. A high abundance of O {\small VI} both in active (super bubble region) and inactive regions of the LMC was found. The dominant regions in the LMC in terms of O {\small VI} abundance are 30 Doradus, SNR 1987 A, N11 and Shapley Constellation III where O {\small VI} is probed upto a small scale variation of $\sim$10 pc. The abundance and properties of O {\small VI} absorption in the LMC are similar to that of the MW and the SMC despite their difference in the metalicity. The correlation of O {\small VI} with the ISM morphologies were studied and found that O {\small VI} absorption in the LMC neither correlates with H$\alpha$ (warm gas) nor with X-ray (hot gas) observations. Surprisingly, O {\small VI} column density shows correlation with X-ray brightnesses in the 30 Doradus region, the most active star formation region in the LMC and it decreases with an increase in the angular distance from the star cluster R136 as the strength of the stellar wind decreases outwards.




\end{document}